# Engineering Perovskite Emissions via Optical Quasi-Bound-States-in-the-Continuum

Evelin Csányi[1,2], Yan Liu[1], Soroosh Daqiqeh Rezaei[3], Henry Yit Loong Lee[1], Febiana Tjiptoharsono[1], Zackaria Mahfoud[1], Sergey Gorelik[4], Xiaofei Zhao[5], Li Jun Lim[5], Di Zhu[1], Jing Wu[1], Kuan Eng Johnson Goh[1,6,7], Weibo Gao[6], Zhi-Kuang Tan[5], Graham Leggett[2], Cheng-Wei Qiu[8,*], and Zhaogang Dong[1,9*]

[1]Institute of Materials Research and Engineering (IMRE), Agency for Science, Technology and Research (A*STAR), 2 Fusionopolis Way, Innovis #08-03, Singapore 138634, Republic of Singapore

[2]Department of Chemistry, University of Sheffield, Brook Hill, Sheffield S3 7HF, United Kingdom

[3]Pennsylvania State University, State College, PA 16801, United States

[4]Singapore Institute of Food and Biotechnology Innovation, Agency for Science, Technology and Research (A*STAR), 31 Biopolis Way, #01-02 Nanos, 138669, Singapore

[5]Department of Chemistry, 3 Science Drive 3, National University of Singapore, 117543, Singapore

[6]Division of Physics and Applied Physics, School of Physical and Mathematical Sciences, Nanyang Technological University, Singapore 637371, Singapore

[7]Department of Physics, National University of Singapore, Singapore 117551, Singapore

[8]Department of Electrical and Computer Engineering, National University of Singapore, 4 Engineering Drive 3, 117583, Singapore

[9]Department of Materials Science and Engineering, National University of Singapore, 9 Engineering Drive 1, 117575, Singapore

*Correspondence and requests for materials should be addressed to Z. D. (email: dongz@imre.a-star.edu.sg) and C.-W. Q. (email: chengwei.qiu@nus.edu.sg).

**ABSTRACT**

Metal halide perovskite quantum dots (PQDs) have emerged as promising materials due to their exceptional photoluminescence (PL) properties. A wide range of applications could benefit from adjustable luminescence properties, while preserving the physical and chemical properties of the PQDs. Therefore, post-synthesis engineering has gained attention recently, involving the use of ion-exchange or external stimuli, such as extreme pressure, magnetic and electric fields. Nevertheless, these methods typically suffer from spectrum broadening, intensity quenching or yield multiple bands. Alternatively, photonic antennas can modify the radiative decay channel of perovskites *via* the Purcell effect, with the largest wavelength shift being 8 nm to date, at an expense of 5-fold intensity loss. Here, we present an optical nanoantenna array with polarization-controlled quasi-bound-states-in-the-continuum (q-BIC) resonances, which can engineer and shift the photoluminescence wavelength over a ~39 nm range and confers a 21-fold emission enhancement of $FAPbI_3$ perovskite QDs. The spectrum is engineered in a non-invasive manner *via* lithographically defined antennas and the pump laser polarization at ambient conditions. Our research provides a path towards advanced optoelectronic devices, such as spectrally tailored quantum emitters and lasers.

**Introduction**

Metal halide perovskites are fast emerging as the next-generation semiconductor materials.[1–5] The explosion of interest in these materials has been driven by their outstanding photoluminescence quantum yield (PLQY) of > 90%, high charge carrier mobility, micrometer scale exciton diffusion lengths and low-cost synthesis.[6–8] Furthermore, by tailoring the chemical composition, size or shape of perovskite nanoparticles, it is possible to produce materials with bandgaps that span a wide spectral range, from ultraviolet to near-infrared regions.[9,10] However, the nature of such bottom-up methods inevitably leads to some degree of non-uniformity in the particle morphology, which can result in line broadening and inconsistent emission wavelengths due to the quantum confinement effect.[11] Furthermore, after the chemical reaction is completed, the luminescence properties of the perovskite are fixed and cannot be fine-tuned *in situ*. Such drawbacks are detrimental in applications that require the refined tuning of emissions at specific wavelengths, such as light-emitting diodes (LEDs),[12–15] photodetectors[16–18] single-photon sources[19] and lasers.[20,21]

Recognizing these challenges, post-synthesis engineering of the perovskite emission properties has gained attention recently. For example, adjustment of the photoluminescence (PL) wavelength has been achieved via ion exchange,[22–24] lattice distortions or phase transitions induced by applying external stimuli such as changes in temperature,[25] pressure[26,27] or strain.[28] Some of these methods allow a PL shift over a broad range; for example, up to ~760 nm (~921 meV) under an extremely high pressure of ~40,000 atmosphere.[27] However, it is typically accompanied by significant broadening of full-width-at-half-maximum (FWHM), intensity quenching and requires a sophisticated setup which is impractical for many applications. Alternatively, leveraging the Rashba,[29,30] Zeeman[31] or Stark effects[32] under DC magnetic field or circularly polarized pumping enables the breaking and detection of the triplet state to manipulate their radiative channels, where a PL shift of ~6.6 nm (~30 meV)[33] can manifest. Nevertheless, these approaches often yield multiple emission with bands, rendering them unsuitable for devices that require a single emission peak. The PL peak shift achieved by various methods listed above was summarized in Table S1 in the supplementary information.

An alternate approach is to use lithographically-defined photonic antennas to modulate the emission properties of these materials.[34–41] By coupling the optical modes of an antenna system to the excitonic states of emitters, the optical response of perovskites can be tailored to achieve unique effects, such as enhanced luminescence,[42,43] optical encoding,[41,44] polarization tuning[45,46] and highly tunable emission chirality.[47,48] However, despite these advancements and the attraction of such an approach, the maximum spectral shift reported to date is only 8 nm (~20 meV) *via* coupling to a microcavity, at the expense of a ~5-fold reduction of the PL intensity.[49]

In this paper, we present an optical nanoantenna array with polarization-controlled quasi bound-states-in-the-continuum (q-BIC) resonances capable of engineering and shifting the PL emission of formamidinium lead iodide ($FAPbI_3$) perovskite quantum dots (QDs) by ~39 nm. The radiative decay channel of the QDs was modified *via* the Purcell effect. As $F_p \propto Q/V_{eff}$, where $F_p$ is the Purcell factor, $Q$ is the quality factor (Q-factor) of the resonance and $V_{eff}$ is the effective mode volume, the use of q-BIC resonances with relatively high Q-factors and compact mode volumes can realize a high $F_p$.[50,51] The coupling was accomplished by the use of top-down lithography to engineer the optical resonances of nanoantennas, without affecting the QDs' physical or chemical properties. Consequently, we demonstrate a homogeneous tuning of the perovskite emission wavelength over wide bandwidth of ~39 nm (~82 meV). Furthermore, the antenna array allows for the switching of the PL wavelength by varying the polarization state of the incident laser. Besides tunability, a PL enhancement of ~21-fold and narrowing of the linewidth was realized. These results provide a non-destructive approach to fine-tune the emission wavelength, saturation and brightness of the perovskite emission at ambient conditions. Incorporation of this design concept could expand the range of applications of perovskites and create new opportunities in the fields of tunable optoelectronics, lighting and display, lasing, and spectrally tailored single-photon emitters.[52–57]

**Results and discussion**

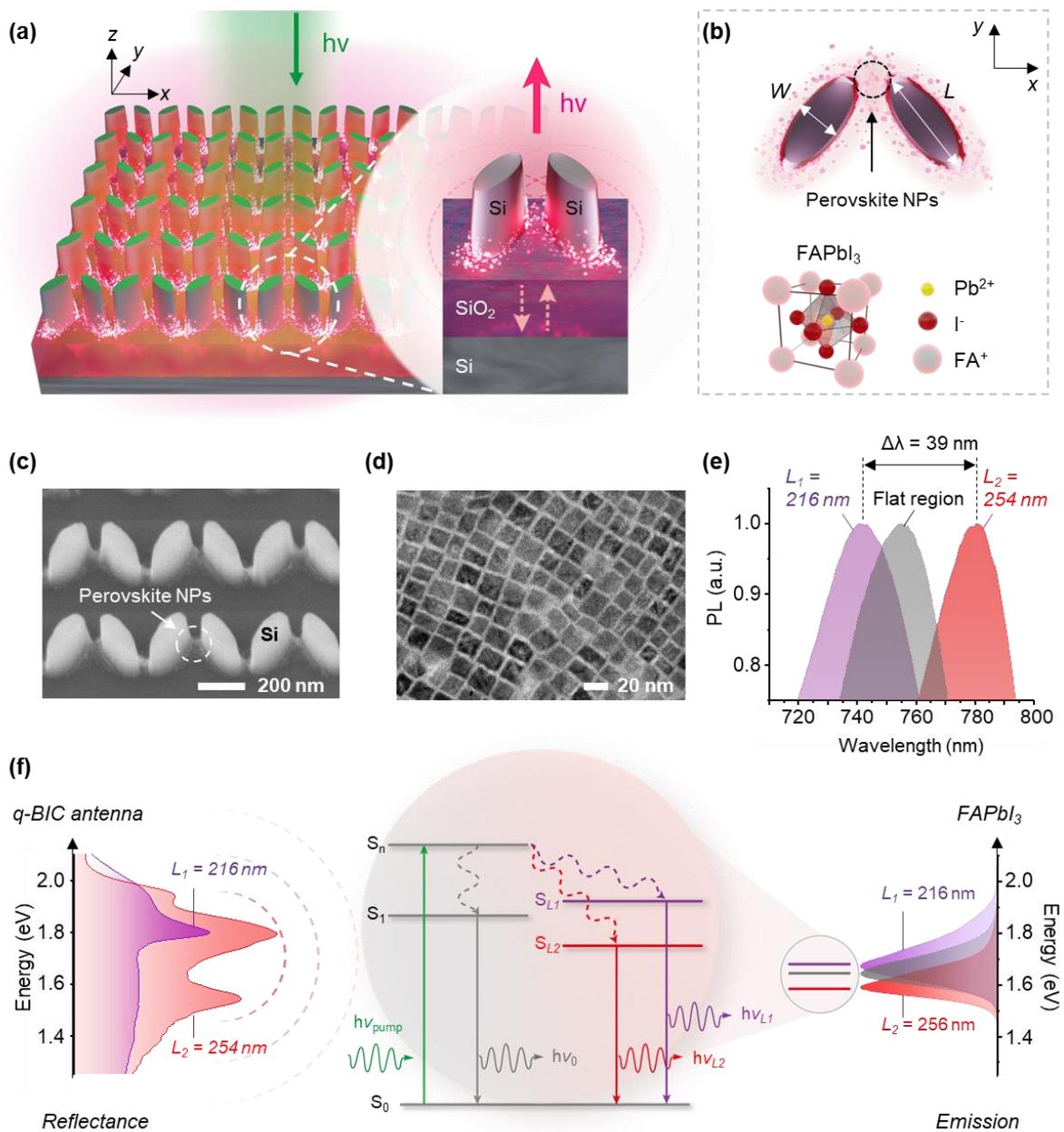

**Figure 1. Design of quasi bound-states-in-the-continuum (q-BIC) resonance for engineering FAPbI$_3$ QD emissions.** (a) Illustration of the silicon (Si) nanoellipse pair array, fabricated on top of a 300-nm-thick SiO$_2$ cavity on Si substrate and coated with FAPbI$_3$ QDs. (b) Top-down view of the nanoellipse pair, with the width (*W*) and length (*L*) annotated and an illustrative crystal structure of FAPbI$_3$. (c) Scanning electron micrograph (SEM) image of the perovskite-covered nanoellipse array, taken at a tilt angle of 30⁰. (d) Transmission electron microscope (TEM) image of the FAPbI$_3$ QDs. (e) Normalized PL spectra of FAPbI$_3$ QDs to demonstrate a tunable range of ~39 nm for the PL emission peak, using $L_1$ = 216 nm and $L_2$ = 256 nm nanoantenna arrays. All spectra were measured under the *y*-polarized pump laser excitation conditions. (f) Illustrative energy level diagram that depicts the interaction between the q-BIC resonance and the emission process of the FAPbI$_3$ QDs. The reflectance spectra of the arrays with $L_1$ and $L_2$ are shown on the left, with the corresponding PL emission on the flat region (without nanoantennae) and on the q-BIC antenna arrays shown on the right. The perovskite is pumped with the 532 nm CW laser. $S_0$, $S_1$, $S_n$ denote the ground state, lowest energy excited state and n$^{th}$ excited state of the perovskite, respectively, while $S_{L1}$ and $S_{L2}$ denote the alternate energy levels provided by the Purcell effect ($F_p$) on the q-BIC nanoantenna arrays with $L_1$ and $L_2$.

Figure 1 shows the silicon (Si) antenna design that is able to engineer and shift the emission wavelength of FAPbI$_3$ QDs via q-BIC resonances. It consists of an array of symmetry-broken amorphous Si nanoellipse pairs. The height of the Si nanostructures was fixed at 360 nm, with a pitch of 430 nm along the *x*-axis and 435 nm along the *y*-axis. We varied the antenna geometry, such that the lengths (*L*) of the ellipses varied from 216 nm to 254 nm, with the corresponding widths (*W*) varied from 79 nm to 103 nm. Asymmetry was introduced by the rotation of the ellipse pair along the *x*-axis by 49⁰ towards each other. To enhance the emission of the perovskite into the far-field, a 300-nm-thick SiO$_2$ cavity is incorporated into the design with Si as the backside reflector. Further details of the array parameters can be found in Figure S1. For simplicity, we refer to the different arrays by lengths only. FAPbI$_3$ nanoparticles were synthesized according to the method by Xue J. et al,[58] separated by size using a centrifuge to remove any aggregates and were deposited onto the antenna array by dip-coating method.[59,60]

The emission of the particles was recorded by mapping the PL signal on the sample surface at ambient conditions using a confocal microscope equipped with a linearly polarized 532 nm continuous wave (CW) laser. Each region was mapped by setting the polarization state of the pump laser alternately to *x*-polarized (with the electric field oscillating along the *x*-axis ellipse pair) and *y*-polarized (along the *y*-axis). The order of *x*- and *y*-polarized PL mapping was randomized to ensure that changes in the PL response were not due to laser-induced degradation of the perovskite. The mean peak position, intensity and FWHM were determined by making measurements at three different points on the sample. As a reference, the PL emission of $FAPbI_3$ QDs was recorded on the flat region of the substrate, for which the reflectance spectrum can be found in Figure S2. The emission band was centered at ~755 nm, close to the values typically reported for $FAPbI_3$ nanocubes,[61] and is within the range of the q-BIC resonance energies of the arrays under study. Figure 1(b) shows the scanning electron micrograph (SEM) image of the nanoantenna with $FAPbI_3$ nanoparticles, with the profile of the perovskite shown by high resolution transmission electron microscope (TEM) image on Figure 1(d). $FAPbI_3$ forms nanocubes with an average lateral size of 10–15 nm in length (see size distribution in Figure S3), enabling a good fit within the gaps between the Si structures, where the near-field effect is highest. The SEM imaging revealed that perovskite nanoparticles preferentially attach to and settle around the nanoellipse structures.

Figure 1(e) presents the range of the perovskite emission wavelength shift achieved under *y*-polarized laser conditions, where the emission at 755 nm on the flat region is also shown for reference. This wide-ranging spectral control was enabled by coupling the excitonic states of the perovskite with the optical modes of the array – the latter of which is directly managed by tailoring the antenna geometry during the lithography process. For instance, when the perovskite is deposited on the array with $L = 216$ nm, the spectrum displays a blue-shifted PL emission at 741 nm, whereas on the array with larger nanoellipse size, $L = 256$ nm, the peak position redshifts to 780 nm. Therefore, the emission wavelength can be shifted to both higher and lower energies according to the energy of the q-BIC resonance the perovskite interacts with. The corresponding reflectance spectra of the arrays and the process which leads to the modified emission properties is illustrated in Figure 1(e).

On the left panel of Figure 1(f), the experimental reflectance spectrum of the short array ($L = 216$ nm) array is presented, which exhibits a single q-BIC mode at 1.80 eV.[39] In contrast, the array with longer ellipses ($L = 256$ nm) reveals two distinct peaks at 1.54 eV and 1.79 eV that are attributed as q-BIC resonances. The appearance of the optical mode at the lower energy has a significant effect on the resulting PL response of the perovskite, which is demonstrated on the right panel. In the case of the $L = 216$ nm array, the PL emission shifts to 1.67 eV towards the high energy q-BIC resonance. On the other hand, when using the $L = 256$ nm array, there are two potential decay channels provided by the two q-BIC modes that exist within the system. Significantly, the perovskite preferentially couples to the lower energy resonance, thus leading to single emission band at 1.59 eV, closer to the low energy q-BIC resonance. The observed spectral shift of the emission wavelength originates from the Purcell effect ($F_p$), which refers to the coupling of the resonant q-BIC mode with the excitonic state of the perovskite. The interaction modifies the spontaneous emission rate of the system, creating an alternate radiative decay channel through the coupled state, where the PL emission wavelength shifts towards the resonance frequency of the coupled mode.[49,62]

**Purcell enhancement of FAPbI$_3$ emission *via* q-BIC resonances**

The coupling of an emitter to a resonant mode will result in a modified radiative decay rate, and the degree of radiation enhancement can be quantified by the Purcell factor ($F_p$)[63]:

$$F_p = \frac{3}{4\pi^2} \frac{Q}{V_{eff}} \left(\frac{\lambda}{n}\right)^3, \tag{1}$$

where $Q$ is the quality factor of the resonance, $V_{eff}$ is the mode volume, $\lambda$ is the emission wavelength and $n$ is the effective mode index.

First, we investigate the Purcell effect of the nanoantennas with $L = 232$ nm, where the resonant mode overlaps spectrally with the excitonic perovskite emission at ~755 nm. To achieve this, the measurements were performed under *y*-polarized conditions, as the symmetry-broken element of the nanoellipse design gives rise to a higher energy q-BIC mode under *x*-polarized light - leading to spectral mismatch and inefficient coupling (Figure S4). The polarization-dependence of the perovskite PL is discussed in detail later on. The resonant mode was analyzed by angle-resolved reflectance, as measured using a back-focal-plane (BFP) microscope setup under *y*-polarized conditions and is presented on Figure 2(a).

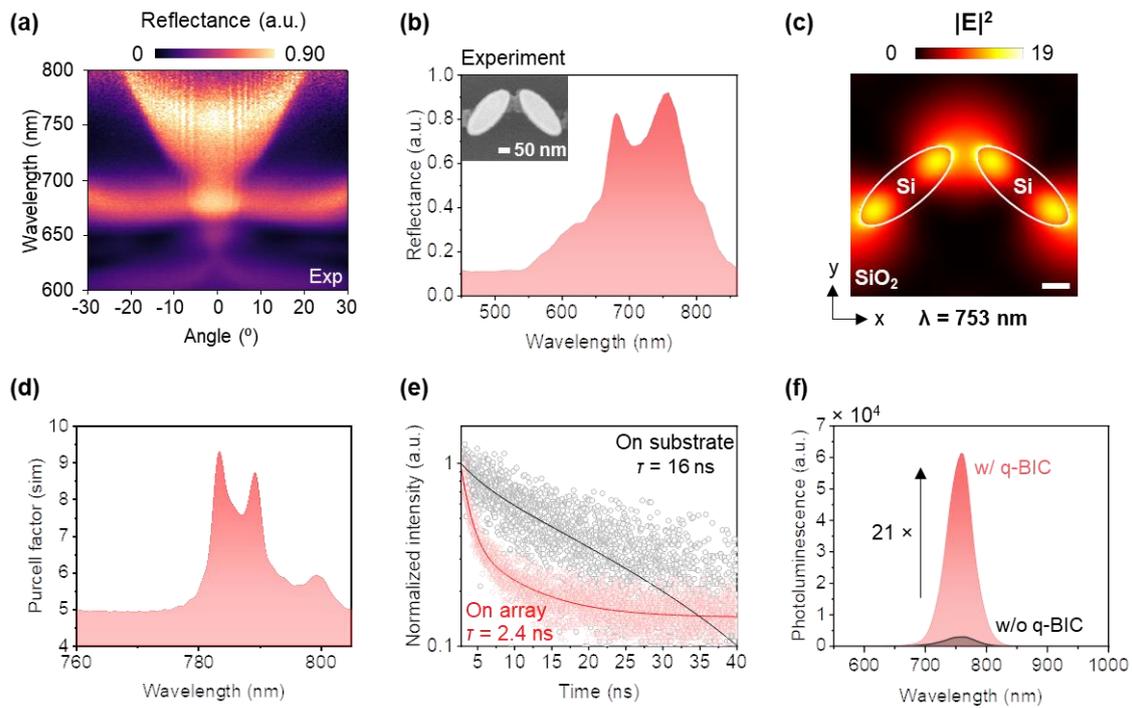

**Figure 2. Purcell enhancement effect of the q-BIC antenna under *y*-polarization.** (a) Back-focal-plane (BFP) image showing the experimental angular-resolved reflectance of an array with $L = 232$ nm and $W = 93$ nm under *y*-polarized incident condition and (b) corresponding experimental reflectance spectra, taken from the cross-section from (a) at an incident angle of 0⁰. The inset shows an SEM image of the perovskite coated nanoantennas. (c) Spatial distribution of the electric field magnitude ($|E|^2$) at $\lambda = 753$ nm for the array, obtained from FDTD simulations. The scale bar corresponds to 100 nm. (d) The corresponding simulated Purcell enhancement factor of the same array as a function of wavelength. (e) Experimental radiative decay rates of FAPbI$_3$ QDs recorded on the flat region of the sample and on an array with $L = 230$ nm, and (f) the PL peak intensity of the perovskite on the same array compared to the PL response on a nanodisk array without q-BIC resonances.

The experimental data shows a spectral band located at 753 nm at the Γ point, attributed to a q-BIC mode.[40] The angle-resolved data was found to agree well with the simulation obtained using finite-difference-time-domain (FDTD) calculations, shown in Figure S5, although the broadening of the experimental spectra is noted. The experimental spectrum at 0⁰ incident angle is presented on Figure 2(b), while electric field distribution of the mode profile at $\lambda = 753$ nm is shown in Figure 2(c), which demonstrates a significant field enhancement effect localized within and around the tips of the nanoellipses. The multipolar decomposition of the scattering cross section revealed that a strong magnetic dipole is manifested in the decomposition around the q-BIC peak, in agreement with the mode pattern distribution (Figure S6). To investigate the Purcell effect of the q-BIC nanoarrays, $F_p$ was calculated using FDTD by placing a point dipole source at the location of the highest field enhancement of the nanoellipse tip, which was found to be at the Si ellipse interface, 5 nm above the surface of the SiO$_2$ layer. The dipole emission wavelength was set to 755 nm with a 100 nm span, and $F_p$ was obtained by averaging the results from the orthogonal dipole orientations. Figure 2(d) shows that the Purcell factor increases sharply to a value of 9.3 at 783 nm, with another sharp peak present at 789 nm that has a value of 8.7. A small, broader feature can also be seen at 799 nm, with a Purcell factor of 5.9. The multiple peaks indicate that the emitter can couple into multiple optical modes of the nanostructure, with the highest enhancement of the radiative decay rate at the frequency closest to the lowest energy q-BIC resonance of the array, where the coupling interaction is the most efficient. According to Fermi's Golden Rule, $F_p$ is directly related to the spectrally integrated Purcell enhancement ratio of $\gamma/\gamma_{res}$, where $\gamma$ and $\gamma_{res}$ are the radiative decay rates of the emitter on the sample area without q-BIC resonances and within the resonant cavity (*i.e.*, the near-field region of the nanoantenna), respectively.[64]

To confirm the theoretical results, we performed PL lifetime measurements to obtain $\gamma_{res}$ and $\gamma$, as shown on Figure 2(e). The perovskite nanoparticles were pumped using a 640 nm *y*-polarized pulsed femto-second (fs) laser at room temperature and the PL decay rate was determined by fitting the experimental data with an exponential decay function. Further details can be found in Figure S7. The PL decay rate of FAPbI$_3$ on the flat region was determined to be 16.1 ns, close to the lifetime of the particles in toluene, and is ascribed to a combination of radiative and nonradiative exciton recombination mechanisms at this temperature.[61] Furthermore, the nanosecond scale suggests the presence of free excitons.[65] To measure $\gamma_{res}$, arrays were fabricated with nanoellipse lengths varied from $L = 216$ nm to $L = 245$ nm. It was found that the rate of spontaneous emission is enhanced on all arrays, with a reduced lifetime within the range of 2-3 ns, without correlation to the array geometry. The array which exhibits the same optical properties as the one discussed earlier (here, $L = 230$ nm) induced a 6.7-fold decrease of the radiative decay rate relative to the flat region, slightly lower than the theoretically calculated value. This enhancement is ~2.3-fold larger than that achieved using hyperbolic metamaterials.[66] Accordingly, a significant increase of the PL brightness was observed within the antenna region. We note that the detected PL intensity is influenced by the concentration of emitters within the sampling area, which may differ on the flat region and on the nanoarrays. Therefore, to evaluate the PL enhancement originating from the Purcell effect, we fabricated nanodisk arrays that provide a similar platform for the perovskite distribution without exhibiting resonant modes in the spectral region of interest. The reflectance spectrum of the array and the SEM image of the perovskite covered surface is presented in Figure S9. The perovskite PL band on the non-resonant array was centered at 758 nm, close to that on the flat region. The nanoarrays promoted a 21-fold enhancement of the PL intensity under *y*-polarized excitation, relative to the arrays without q-BIC resonances, as shown on Figure 2(f). This enhancement is nearly twice of the value obtained from the Purcell factor calculations, meaning that there are additional enhancement effects, such as the enhanced pump laser absorption and enhanced directionality due to the q-BIC nanoantenna array.[67] Furthermore, the Purcell effect promotes a narrowing of the spectral FWHM linewidth from 63 nm to 49 nm as shown in Figure S10, resulting from the more homogeneous emission at the resonant wavelength. These findings therefore demonstrate the value of nanoantennas to achieve a non-invasive modification of the radiative decay rate of perovskites, enabling a significant improvement of the PL brightness.

**Tunable PL emission via geometrical engineering of the array**

Next, we investigated the degree of emission wavelength shift enabled by the engineering of the array geometry. The emission response of the perovskite is modulated by the coupled state with the q-BIC mode, which is directly related to the size of the nanoellipses. Figure 3(a-b) presents the evolution of the reflectance spectrum of arrays with increasing ellipse size under *x*- and *y*-polarized conditions. As expected, the increase in nanostructure size redshifts the q-BIC resonance wavelength. Furthermore, under *y*-polarized conditions, the resonance becomes more prominent and higher order modes emerge. The dashed line on the graphs represents the typical spectral position of the PL band of FAPbI$_3$ on the flat region.

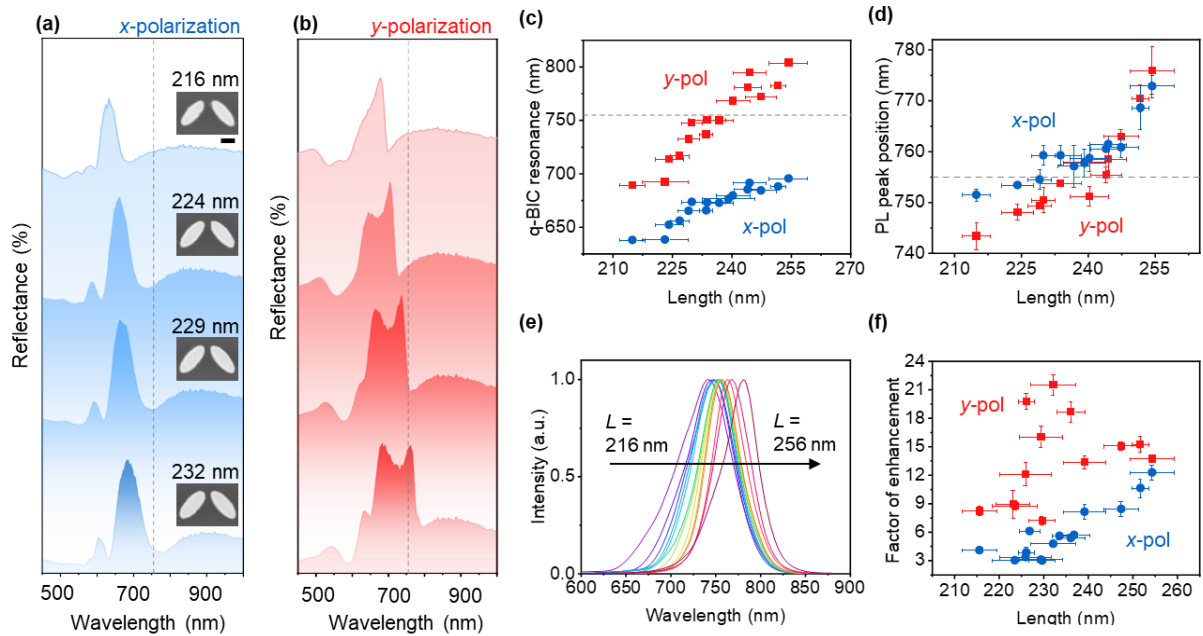

**Figure 3. Control of the PL shift and brightness of perovskites via geometrical engineering of the array.** Reflectance spectra of nanoarrays with different nanostructure sizes, with the corresponding SEM images of the nanoellipse pairs shown in the inset under (a) *x*-polarized and (b) *y*-polarized incident light. The dashed line represents the PL peak position of the perovskite on the flat area of the substrate on all graphs. (c) Spectral position of the q-BIC peak under *x*- and *y*-polarized conditions as a function of the nanoellipse lengths. For the *y*-polarized data, the lowest energy q-BIC peak positions are presented. (d) PL emission wavelength of the perovskite on arrays with varying nanostructure lengths, using *x*- and *y*-polarized laser excitation. (e) Demonstration of the PL band shift from the smallest geometry arrays ($L = 216$ nm) to the largest ones ($L = 254$ nm) under *y*-polarization. (f) Factor of enhancement of the PL intensity relative to that on the nanodisk arrays without q-BIC resonances, as a function of the nanoellipse size.

Figure 3(c) summarizes the size-dependent redshift of the q-BIC energy under both polarization states. Here, the wavelength of the lowest energy peak is presented from the y-polarized reflectance data. When $L = 216$ nm is increased to 256 nm, the optical mode redshifts from 690 nm to 804 nm under y-polarization, and from 638 nm to 696 nm under x-polarization. The polarization switching results in a significant difference in the range of 50-110 nm of the q-BIC spectral positions. Following the principles of dyadic Green's function[64], the perovskite emission characteristics change according to the properties of the resonant mode that the excitonic states couple with. The relationship between the Purcell factor and the emission wavelength shift was derived in detail in the work of H. P. Adl et al.[66] Consequently, the emission wavelength redshifts with increasing nanoellipse size and exhibits a polarization-dependent PL band position. This phenomenon is shown in Figure 3(d), with the representative normalized PL spectra for nanoellipse lengths ranging from 216 nm to 256 nm under y-polarized illumination displayed on Figure 3(e). The maximum wavelength shift range achieved by adjustment of the geometry was ~39 nm under y-polarized excitation and a shorter range of ~19 nm under x-polarization.

In addition, we have found that the nanoellipse size has a significant influence on the polarization-dependent PL emission, where smaller nanostructures, *i.e.*, higher energy q-BIC modes, result in larger differences. Here, the y-polarization induced q-BIC frequency is still close enough in energy for the perovskite excitons to couple to, thus introducing the alternate radiative decay channel closer to the coupled state. Moreover, the degree of spectral overlap between the excitonic emission wavelength and the resonant mode determines the Purcell enhancement effect of the array. Figure 3(f) presents the factor of enhancement of arrays with varying nanostructure sizes, illustrating that as the q-BIC resonance gets closer to full spectral overlap, the factor of enhancement increases up to ~21-fold under y-polarization ($L = $ ~226-236 nm). Further increase in the structure size results in a spectral mismatch which in turn decreases the enhancement factor to ~15. On the other hand, using *x*-polarized excitation on the same arrays, the achieved enhancement is only ~6-fold. However, this value increases continuously with increasing nanoellipse size owing to the more efficient coupling. Simultaneously, the emission linewidth is reduced, as shown in Figure S10, where the PL linewidth narrows by up to ~27% under *y*-polarization and ~19% for *x*-polarization. Lastly, we note that arrays were designed and fabricated in which the nanoellipse rotational angles were varied to achieve near-parallel orientations relative to each other, in order to improve the Q-factor of the resonances.[39] Although the rotated geometries enabled higher quality q-BIC resonances, we did not observe a significant improvement or trend in the FWHM of the perovskite emission.

**Polarization switchable perovskite PL emission wavelength**

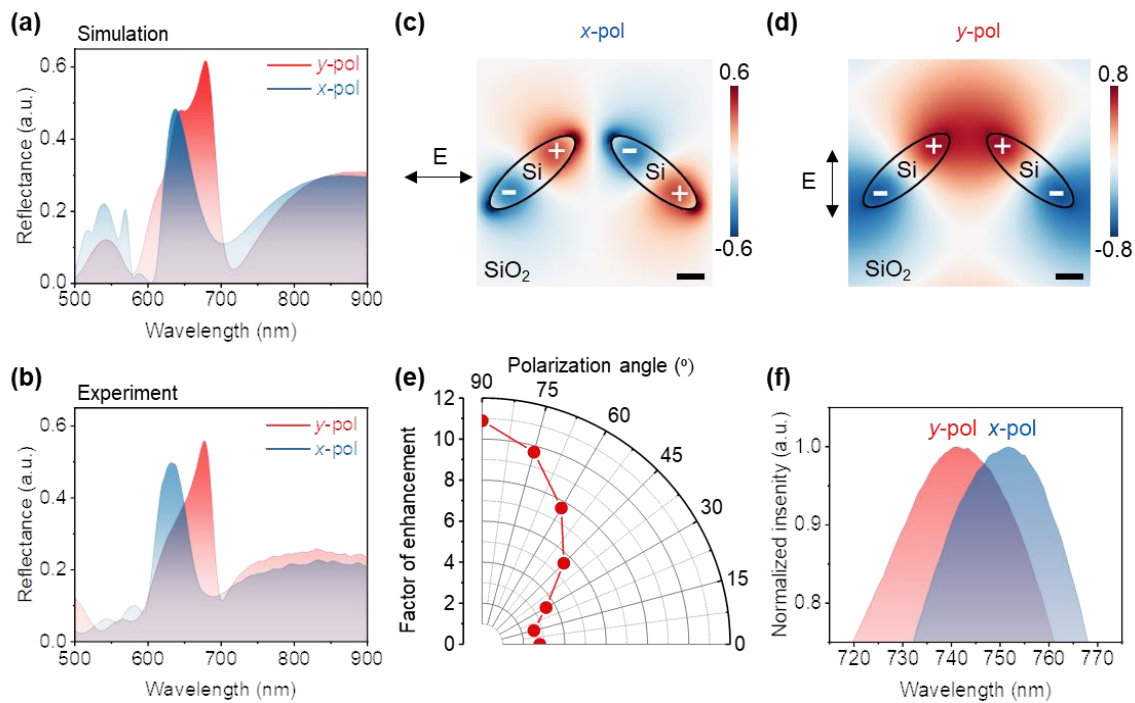

**Figure 4. Polarization switchable perovskite PL emission wavelength.** (a) Simulated and (b) experimental reflectance spectra of a nanoarray with $L = 216$ nm and $W = 79$ nm when illuminated with *x*- and *y*-polarized light. Spatial distribution profile of the $E_z$ component of the resonant modes under (c) *x*- and (d) *y*-polarized conditions. (e) PL enhancement factor as a function of the polarization state of the excitation laser, changing from *x*-polarization (0º) to *y*-polarization (90º). (f) Normalized PL band of the perovskite coated array that demonstrates the emission wavelength shift under *x*- and *y*-polarized laser excitation.

To demonstrate the polarization switching of the emission wavelengths, we selected the nanoarray with the shortest dimensions ($L = 216$ nm), which supports optical modes that are furthest in energy from the perovskite excitonic states. Figure 4(a) and (b) shows the simulated and experimental reflectance spectra of the array when the polarization state of the light source is varied. The q-BIC resonances can be observed at 690 nm for *y*-polarized pump laser, and at 637 nm under *x*-polarized pump condition. The corresponding $E_z$ component of the electric field at the respective resonant wavelengths is shown in Figure 4(c) and Figure 4(d), presenting the distinct mode profiles under the two polarization conditions. Therefore, the emitters in the vicinity of the nanoantennas experience significantly different electromagnetic environments, which directly influences the Purcell effect under the given polarization. This difference can be observed as the polarization state of the incident laser light is varied in increments from *x*-polarization (0º) to *y*-polarization (90º), as shown in the polar plot in Figure 4(e), illustrating the gradual increase in the enhancement factor from 3 to 12. Most notably, the *y*-polarized laser excitation introduces a radiative decay channel that gives rise to an emission band at 741 nm, while switching to *x*-polarization, causes a redshift to 751 nm. The feature of switchable emission wavelengths, as well as the enhanced PL intensity, would be of particular interest in the design of single-photon emitter applications.[68] By trapping individual QD in the gap, one may realize efficient single-photon sources which emission is controlled by the antenna, which could mitigate the problems of spectral diffusion due to inhomogeneities during the material synthesis. In addition, the possibility of designing double resonances at both the pump laser frequency and emission wavelength could further enhance the efficiency of the system.[69]

**Conclusions**

In this paper, we present the use of Si nanostructured optical antennas to engineer the emission of FAPbI$_3$ emission via q-BIC resonances. By exploiting the variability of the q-BIC resonance energy, a tuning range of ~39 nm of the emission wavelength was achieved *via* the Purcell effect. Furthermore, we demonstrate that the q-BIC array facilitates switchable emission wavelengths of up to ~10 nm, as controlled by the polarization state of the pump laser. Meanwhile, maximizing the coupling efficiency between the resonant mode and the perovskite excited state allows enhancement of the PL intensity by a factor of 21-fold. FDTD simulation of the optical properties of the arrays and the Purcell factor provided further insight into the underlying radiative decay rate enhancement, showing good agreement with the experimental results. In future work, further optimization of the antenna design could enable polarization-dependent resonant modes that are further apart in energy to promote a higher separation of the potential radiative decay channels for the perovskite excited state. The presented work could pave the way for interesting optoelectronic applications, including wavelength-tunable single-photon sources and lasers.

**Methods**

**Fabrication of the q-BIC nanoantenna array.** 360-nm-thick amorphous silicon (a-Si) was deposited on top of Si substrates with 300 nm SiO$_2$ overlayer (purchased from Silicon Valley Microelectronics, Inc), using plasma-enhanced chemical vapor deposition (PE-CVD, Oxford Instruments Plasmalab System 100). The process was carried out using SiH$_4$ at a flow rate of 45 sccm and Ar at 30 sccm. The temperature was set at 250 ºC with a pressure of 8.0 mTorr, a RF power of 50 watts and inductively-coupled plasma (ICP) power of 3000 watts. When ready, the sample was spin-coated (5k round-per-minute (rpm)) with 2% hydrogen silsesquioxane (HSQ, Dow Corning XR-1541 E-beam resist), after which the lithography process was carried out immediately. Using electron beam lithography (EBL, Elionix ELS-7000), the patterns were written with an electron acceleration voltage of 100 keV, a beam current of 500 pA, and the exposure dose of 8,960 – 20,480 $\mu C/cm^2$ to achieve elliptical shapes with lengths ranging from 215 nm to 254 nm and widths ranging from 69 to 103 nm. To develop the sample, it was immersed in a solution of NaOH/NaCl (1% wt./4% wt. in de-ionized water) for 60 seconds and deionized water for another 60 seconds, followed by rinsing with acetone, isopropyl alcohol (IPA) and drying with nitrogen.[70] Finally, to create the nanopatterns, the sample was etched by inductively-coupled-plasma (ICP, Oxford Instruments Plasmalab System 100) using $Cl_2$ at a flow rate of 22 sccm. The ICP and RF power were 300 watts and 100 watts, respectively, under 5 mTorr and a temperature of 10 °C.

**Synthesis of formamidinium lead iodide quantum dots ($FAPbI_3$ QD) and sample coating.**
The details of the $FAPbI_3$ QD synthesis can be found in the work by Xue J. et al.[58] The synthesized $FAPbI_3$ solution was centrifuged at 6k rpm for 10 min. The supernatant was separated, and the centrifugation cycle was repeated 3 times. The supernatant was then filtered using a microfilter with a 0.2 µm pore size and diluted 1:2 with toluene to yield the final solution for dip-coating. To deposit the $FAPbI_3$ QDs, the q-BIC arrays were first oxygen plasma treated by exposure to inductively-coupled plasma (ICP, Oxford Plasmalab System 80). A pressure of 10 mTorr, a 100 watts DC power and 500 watts coil power was used during the process with an $O_2$ flow rate of 50 sccm at room temperature. Immediately after, the sample was dip-coated in the $FAPbI_3$ QD solution at a speed of 0.8 mm min$^{-1}$ in a nitrogen filled glove box, at room temperature. The samples were stored in the dark in nitrogen atmosphere at room temperature until further use.

**Spectroscopic characterization.** Reflectance measurements were carried out using a polarized broadband light source on a CRAIC UV-VIS-NIR micro-spectrophotometer equipped with a ×5 objective lens and a numerical aperture of 0.12. The measurements were calibrated using a NISR calibration sample (CRAIC Technologies) to record the absolute reflectance and operated with an integration time of 100 ms with an average of 50 measurements per spectrum. To obtain the angle-resolved reflectance measurements, back-focal-plane (BFP) spectroscopy was carried out with an Andor SR-303i spectrograph and inverted optical microscope (Nikon Ti-U), where a × 50 objective lens was selected with NA of 0.6, using a halogen lamp and EMCCD detector (Andor Newton).

**Photoluminescence (PL) and PL lifetime measurements.** The PL measurements of the perovskite-coated q-BIC arrays were carried out at ambient conditions unless otherwise stated. PL mapping measurements were carried out and the spectra were extracted from the same 3 positions on each array. The mapping was conducted on a WITec Alpha 300 S Scanning Near-field Optical Microscope in confocal microscopy mode. A linearly polarized 532 nm CW laser with a power of 50 μW was focused on the sample through a 50× objective with a NA of 0.6. An integration time of 0.05-0.10 second was used. During the analysis process, the resulting spectra was smoothed using the Savitzky-Golay filter to reduce noise. The time-resolved PL was collected on a Picoquant Microtime 200 TCSPC system coupled with an Olympus microscope in confocal configuration, with a 20× objective lens. The data was recorded and fitted with Picosharp software. A pulsed laser of 640 nm, with pulse duration of 70 $ps$ was used to excite the sample, and the repetition rate was set to 40 MHz, with an average power of 1 μW. To collect the spectrally integrated PL (660-800 nm range), a Si single photon avalanche photodiode (APD) was used. The instrument response function (IRF) was obtained using the scattered excitation light from the sample, from which the instrument response time was determined (~240 $ps$).

**Scanning electron microscopy (SEM) and transmission electron microscopy (TEM) characterizations.** Prior to the SEM imaging, the samples were coated with 2 nm Cr to reduce the charging. The metal was deposited using electron beam evaporation (Denton Explorer E-beam Evaporator), at a deposition rate of 1 Å/s and a pressure of $4 \times 10^{-6}$ Torr. SEM imaging (Elionix, ESM-9000) was carried out using an electron acceleration voltage of 5 kV. The TEM imaging was performed using FEI Titan equipped with a Gatan OneView camera, and was carried out at 200 kV.

**Numerical simulations and Purcell enhancement calculations.** The far-field spectra and the near-field distributions were obtained using three-dimensional finite-difference time-domain (3D-FDTD) simulations with Lumerical FDTD Solutions software. The elliptical structures were placed in a 436 nm × 436 nm × 1500 nm unit cell with a 4 nm × 4 nm × 4 nm mesh around them. The unit cell had perfectly matched layer (PML) in the z direction and periodic boundary conditions in the $x$ and $y$ directions. The reflectance spectrum was recorded at normal incidence condition by a monitor placed above a plane wave source, with the polarization of the incident light wave set to be along the $x$- or $y$-axis. The refractive index ($n$) and extinction coefficient ($k$) values of Si were taken from previous measurements.[70] The optical properties of $SiO_2$ were selected from the software's material database from Palik.[71] To obtain the angle-dependent reflectance, a broadband fixed angle source technique (BFAST) plane wave was used to provide broadband simulations at a range of illumination angles. The Purcell factor calculation was carried out using the built-in calculation method of the software, using the ratio of the dipole emission power over the total emitted power. The domain was limited by the PML boundary conditions. The point dipole source was placed onto a 20×20 array of the nanoellipse pairs, at the area of the highest field enhancement of a nanoellipse in the middle of the array. The emission center was set to 755 nm, with a 100 nm span, and the simulation results were averaged from the orthogonal dipole orientations in the $x$-, $y$- and $z$-directions.

**Author contributions.**

E.C., C.-W.Q. and Z.D. conceived the concept, designed the experiments, wrote, and revised the manuscript. F.T. and Z.D. did the nanofabrication of the nanostructures. E.C. performed optical reflectance measurements, $FAPbI_3$ QD size separation, dip-coating and PL mapping measurements. H.Y.L.L. performed the Cr evaporation and SEM imaging. Z.M. carried out the TEM imaging. S.G. performed the PL lifetime measurements and designed the back-focal plane reflection setup. E.C., Y.L. and S.D.R. performed the finite-difference time-domain (FDTD) simulations. X.Z, L.J.L. and Z.-K.T. prepared and characterized the perovskite QDs. D.Z., J.W., K.E.J.G., W.G., and G.L. participated in discussions and provided insightful suggestions. All authors analyzed the data, read, and corrected the manuscript before the submission.

**Acknowledgements**

This research work is supported by Agency for Science, Technology and Research (A*STAR) under its <AME IRG (Project No. A20E5c0093)>, < Career Development Award grant (Project No. C210112019)>, <MTC IRG (Project No. M21K2c0116 & M22K2c0088)> and Quantum Engineering Programme 2.0 (Award No. NRF2021-QEP2-03-P09 & NRF2022-QEP2-01-P07)>. K.E.J.G. acknowledges funding support from Agency for Science, Technology and Research (#21709), Singapore National Research Foundation Grant CRP21-2018-0001 and Quantum Engineering Programme 2.0 Grant NRF2021-QEP2-02-P07. C.-W.Q. acknowledges financial support from the National Research Foundation, Prime Minister's Office, Singapore under the Competitive Research Programme Award NRF-CRP22-2019-0006.

*Supporting Information*

# Engineering Perovskite Emissions via Optical Quasi-Bound-States-in-the-Continuum

Evelin Csányi[1,2], Yan Liu[1], Soroosh Daqiqeh Rezaei[3], Henry Yit Loong Lee[1], Febiana Tjiptoharsono[1], Zackaria Mahfoud[1], Sergey Gorelik[4], Xiaofei Zhao[5], Li Jun Lim[5], Di Zhu[1], Jing Wu[1], Kuan Eng Johnson Goh[1,6,7], Weibo Gao[6], Zhi-Kuang Tan[5], Graham Leggett[2], Cheng-Wei Qiu[8,*], and Zhaogang Dong[1,9*]

[1]Institute of Materials Research and Engineering (IMRE), Agency for Science, Technology and Research (A*STAR), 2 Fusionopolis Way, Innovis #08-03, Singapore 138634, Republic of Singapore
[2]Department of Chemistry, University of Sheffield, Brook Hill, Sheffield S3 7HF, United Kingdom
[3]Pennsylvania State University, State College, PA 16801, United States
[4]Singapore Institute of Food and Biotechnology Innovation, Agency for Science, Technology and Research (A*STAR), 31 Biopolis Way, #01-02 Nanos, 138669, Singapore
[5]Department of Chemistry, 3 Science Drive 3, National University of Singapore, 117543, Singapore
[6]Division of Physics and Applied Physics, School of Physical and Mathematical Sciences, Nanyang Technological University, Singapore 637371, Singapore
[7]Department of Physics, National University of Singapore, Singapore 117551, Singapore
[8]Department of Electrical and Computer Engineering, National University of Singapore, 4 Engineering Drive 3, 117583, Singapore
[9]Department of Materials Science and Engineering, National University of Singapore, 9 Engineering Drive 1, 117575, Singapore

*Correspondence and requests for materials should be addressed to Z. D. (email: dongz@imre.a-star.edu.sg) and C.-W. Q. (email: chengwei.qiu@nus.edu.sg).

| Method | Photoluminescence wavelength shift (nm)* | Conditions | Reference |
|---|---|---|---|
| **Ion-exchange** | 300 | Exposure to halide salts | [1] |
| **Temperature** | 30 | 12 K to 160 K temperature change | [2] |
| **Pressure** | 760 | ~6,900 atm to ~40,000 atm pressure change | [3] |
| **Strain** | 3 | Suspension over $SiO_2$ rings with biaxial strain | [4] |
| **Magnetic field** | 6.6 | 0 T to 0.47 T change in magnetic field at 293 K | [5] |
| **Rashba effect** | 28 | Reflection and transmission geometry PL setup at 300 K | [6] |
| **Zeeman effect** | 0.7 | 0 T to 16 T change in magnetic field at 4.2 K | [7] |
| **Stark shift** | 2.6 | Use of co- and counter-circularly polarized pump−probe configuration | [8] |

**Table S1.** Photoluminescence (PL) wavelength shift achieved by different methods outlined in the literature, including a brief note on the conditions at which the changes were observed. The values were extracted from the results provided in each work. *As some of these approached yield multiple PL bands due to a splitting phenomenon, we quote the emission wavelength shift of the second peak relative to the main PL peak position.

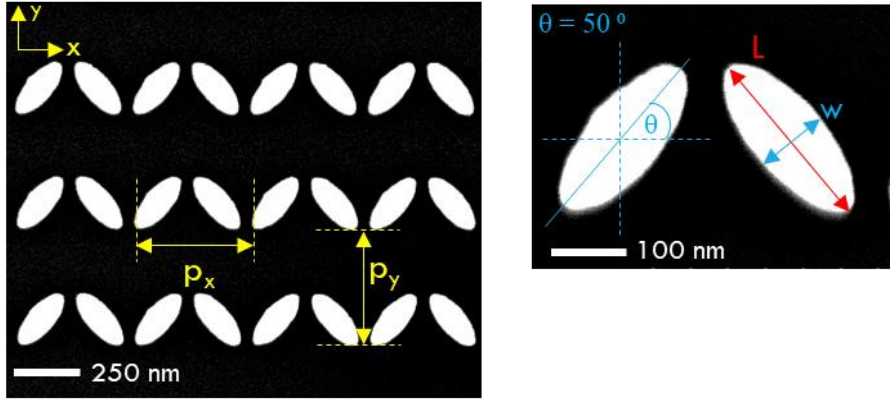

**Figure S1.** Representative SEM images of a nanoellipse array, showing the measured parameters extracted from each array, such as the length of the ellipse (*L*), width (*W*). The pitch in the *x*-direction ($d_x$) and *y*-direction ($d_y$) were fixed at 430 nm and 435 nm, respectively. The rotational angle (θ) was fixed at 49º.

| *L* (nm) | StDev (nm) | *W* (nm) | StDev (nm) |
|---|---|---|---|
| 216 | 4 | 79 | 5 |
| 223 | 6 | 69 | 3 |
| 224 | 5 | 75 | 4 |
| 225 | 6 | 77 | 1 |
| 226 | 2 | 85 | 4 |
| 227 | 2 | 76 | 7 |
| 229 | 5 | 81 | 2 |
| 230 | 3 | 79 | 3 |
| 232 | 6 | 89 | 2 |
| 234 | 2 | 77 | 1 |
| 236 | 3 | 97 | 2 |
| 237 | 4 | 85 | 3 |
| 239 | 7 | 93 | 5 |
| 247 | 4 | 101 | 6 |
| 252 | 2 | 104 | 8 |
| 254 | 5 | 103 | 4 |

**Table S2.** Extracted *L* and corresponding *W* parameters of the nanoellipses from SEM images, including the standard deviation (StDev) within the same array for each value.

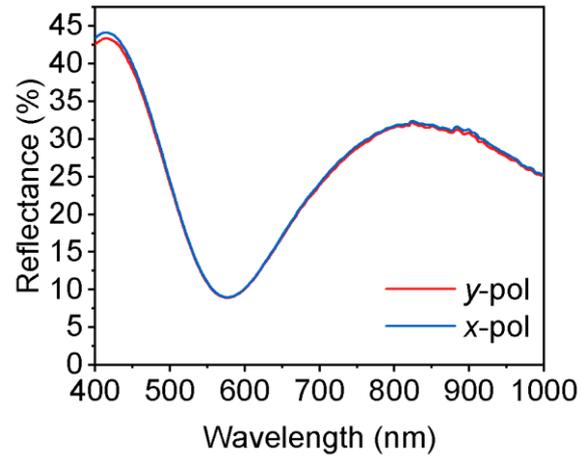

**Figure S2.** Reflectance spectrum of the flat area of the substrate with 300 nm $SiO_2$ overlayer, under *x*- and *y*-polarized light.

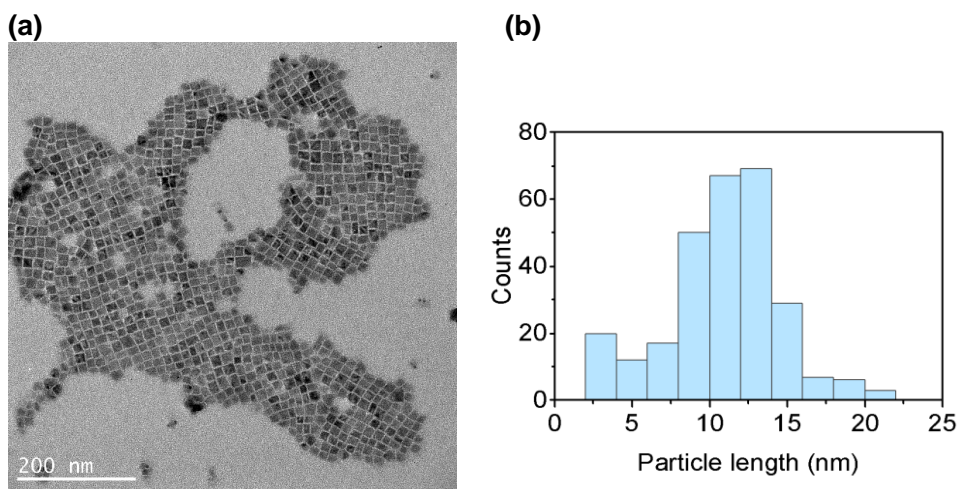

**Figure S3.** (a) Transmission electron microscope (TEM) image of formamidinium lead iodide (FAPbI$_3$) particles and (b) particle size distribution obtained from TEM images.

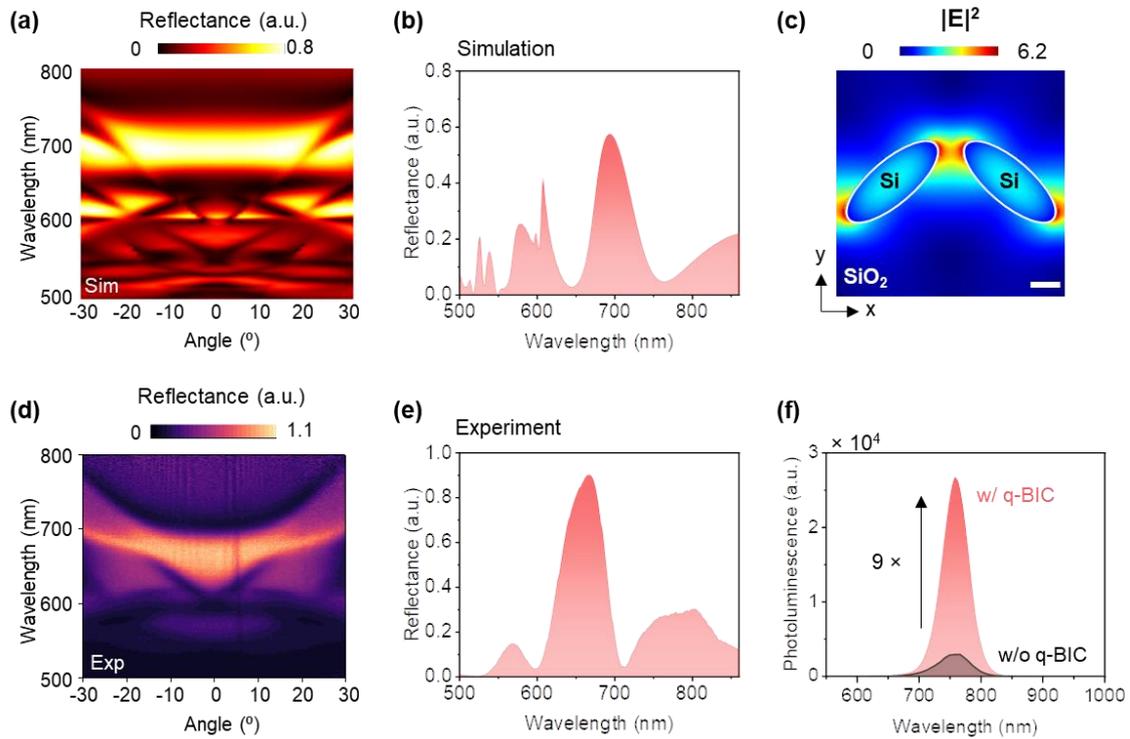

**Figure S4.** (a) Simulated angular-resolved reflectance of the array with $L$ = 232 nm and $W$ = 93 nm under $x$-polarized condition and corresponding. (b) The simulated reflectance spectra at the incident angle of 0⁰. (c) The electric field magnitude distribution ($|E|^2$) at λ = 693 nm for the nanoellipse pair. The scale bar corresponds to 50 nm. (d) Experimental angle-resolved reflectance of the same array with the (e) reflectance spectrum demonstrated at 0⁰ incident angle $x$-polarized conditions. (f) Corresponding PL peaks of the perovskite on the nanoarray and on the reference nanodisk array without q-BIC resonances, excited with $x$-polarized laser.

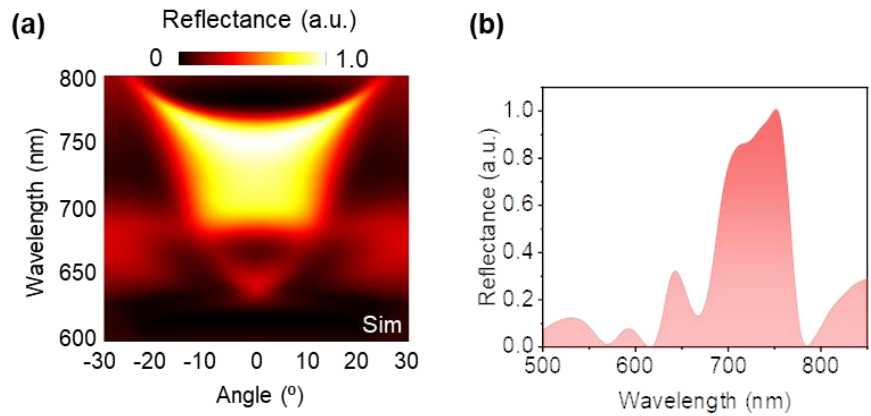

**Figure S5.** (a) Simulated angular-resolved reflectance of an array with $L$ = 232 nm and $W$ = 93 nm under *y*-polarized incident condition and (b) reflectance spectra at the incident angle of 0⁰.

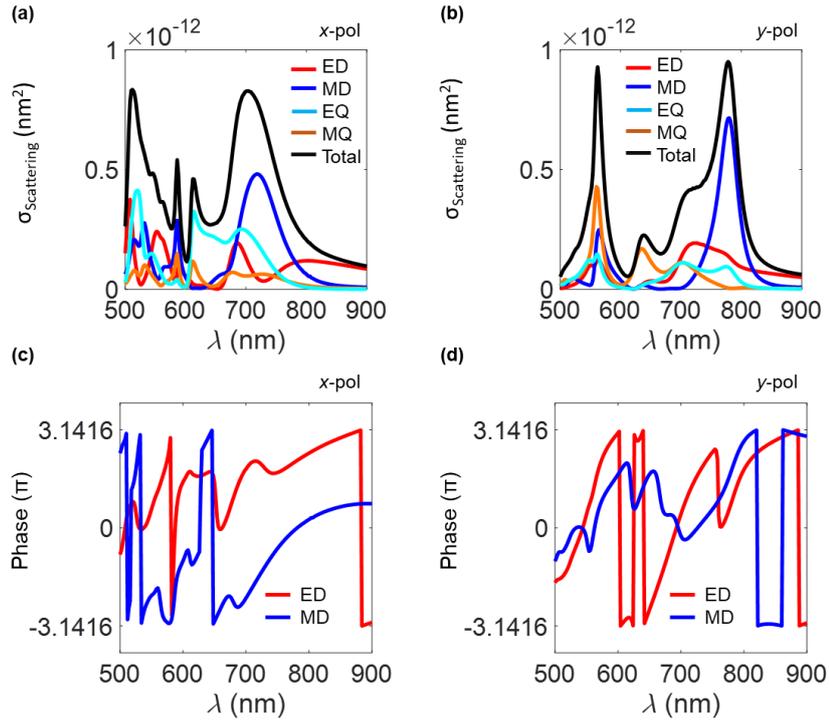

**Figure S6.** Multipolar decomposition of the scattering cross section of the array with *L* = 232 nm and *W* = 93 nm under (a) *x*-polarized and (b) *y*-polarized incident condition, and corresponding phase distribution under (c) *x*-polarized and (d) *y*-polarized light.

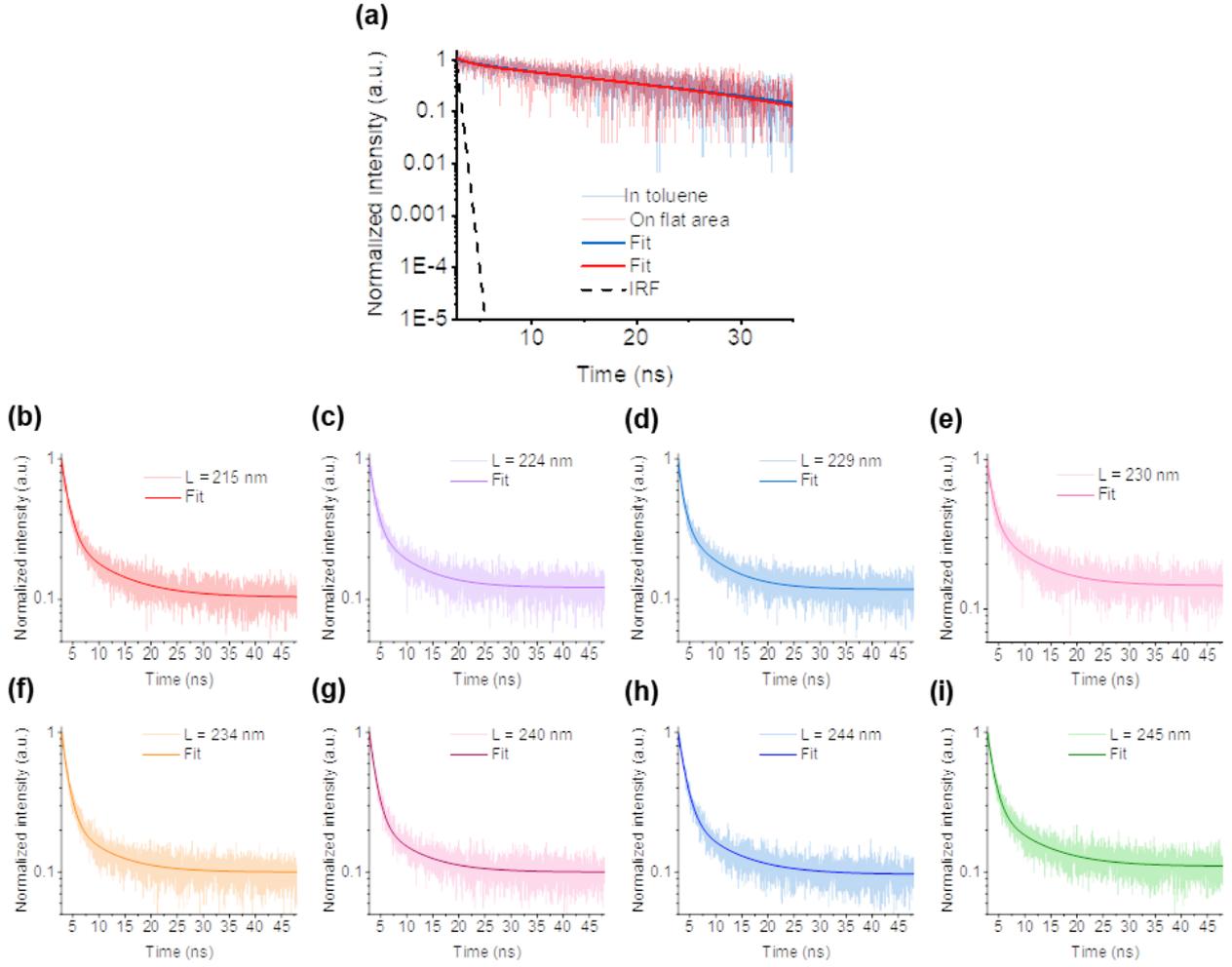

**Figure S7.** PL lifetime decay measurements of FAPbI$_3$ (a) in toluene and on the flat area of the substrate, along with the instrument response function (IRF), for reference. Spectra (b-i) shows the decay curves recorded on arrays with lengths varying from $L = 216$ nm to $L = 245$ nm. All curves were with fitted the following exponential decay function:

$$y = A_1 e^{(-\frac{x}{t_1})} + A_2 e^{(-\frac{x}{t_2})} + y_0$$

Where, $A_1$ and $A_2$ are amplitudes, $t_1$ and $t_2$ are time constants and $y_0$ is the offset. The average lifetime ($\tau$) was obtained via the following:

$$\tau = \frac{A_1 t_1 + A_2 t_2}{A_1 + A_2}$$

Table S3 summarises the components obtained from the fitted data presented above and Figure S8 shows the corresponding reflectance spectra of the arrays.

**Table S3.** Components obtained from the lifetime decay curve fitting presented in Figure S4.

|  | Lifetime (τ, ns) | $A_1$ | $t_{1\,(ns)}$ | $A_2$ | $t_{2\,(ns)}$ | $y_0$ |
|---|---|---|---|---|---|---|
| IRF | 0.24 | 46738.69 | 0.24 | - | - | 22.76 |
| In toluene | 12.68 | 1.36 | 1.65 | 1.07 | 26.70 | -0.16 |
| On the flat area | 17.64 | 0.50 | 3.14 | 1.00 | 24.91 | -0.10 |
| $L$ = 216 nm | 2.63 | 48.19 | 7.33 | 153.71 | 1.16 | 28.14 |
| $L$ = 224 nm | 2.75 | 48.00 | 7.56 | 162.27 | 1.33 | 28.27 |
| $L$ = 229 nm | 2.11 | 46.91 | 5.91 | 165.20 | 1.03 | 26.98 |
| $L$ = 230 nm | 2.42 | 40.14 | 6.31 | 123.75 | 1.16 | 26.96 |
| $L$ = 234 nm | 3.31 | 43.43 | 7.14 | 76.24 | 1.12 | 27.38 |
| $L$ = 240 nm | 2.83 | 47.74 | 6.83 | 108.86 | 1.08 | 27.77 |
| $L$ = 244 nm | 2.98 | 106.37 | 1.26 | 41.36 | 7.40 | 27.96 |
| $L$ = 245 nm | 3.09 | 144.47 | 1.38 | 46.54 | 8.39 | 28.63 |

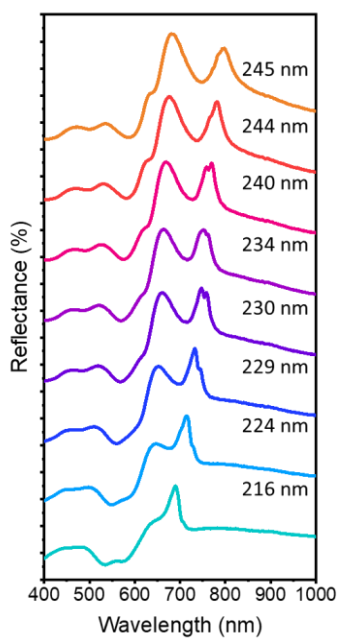

**Figure S8.** Experimental reflectance spectra of the arrays from the time-resolved PL study.

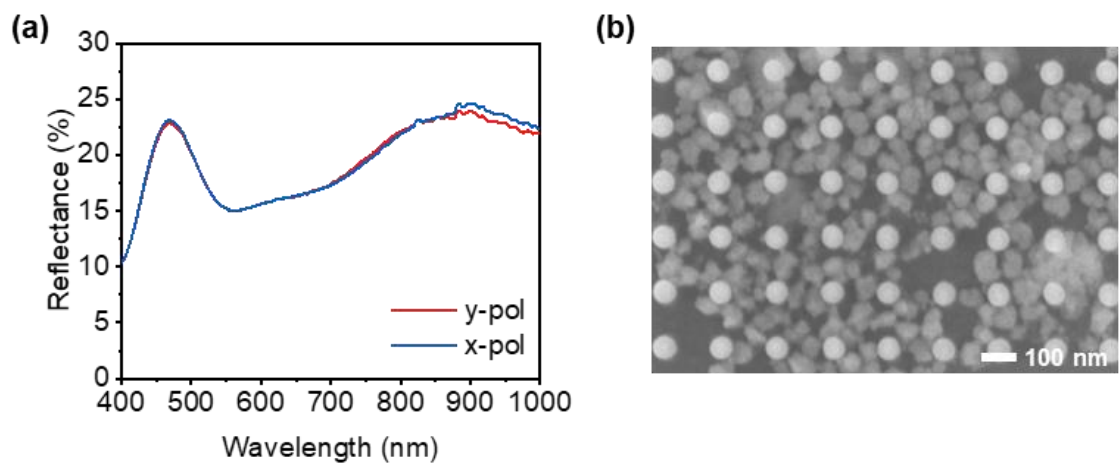

**Figure S9.** (a) Reflectance spectrum of the nanodisk array without optical modes, fabricated as a reference to investigate the PL enhancement effect of the q-BIC resonant nanoarrays with q-BIC resonances. (b) SEM image of the nanodisk array coated with the perovskite nanoparticles.

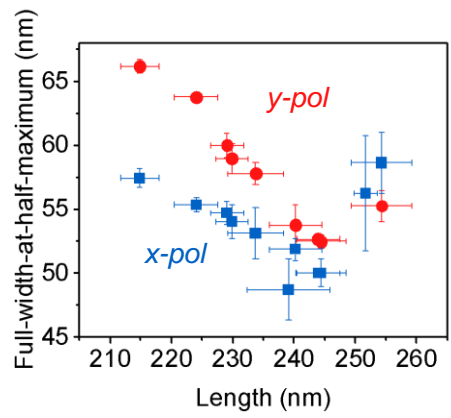

**Figure S10.** Full-width-at-half-maximum (FWHM) of the PL emission peak on nanoantennas under both *x*- and *y*-polarized conditions as a function of the nanoellipse lengths.